\documentclass[
]{ceurart}

\usepackage{listings}
\usepackage{hyperref}
\usepackage{url}

\begin{document}

\copyrightyear{2022}
\copyrightclause{Copyright for this paper by its authors.
  Use permitted under Creative Commons License Attribution 4.0
  International (CC BY 4.0).}

\conference{IAIL 2024 - 3rd International Workshop on Imagining the AI Landscape After the AI Act (In conjunction with The Third International Conference on Hybrid Human-Artificial Intelligence), Malmö, Sweden, June 10, 2024}

\title{“My Kind of Woman”: Analysing Gender Stereotypes in AI through The Averageness Theory and EU Law}

\author[1,2]{Doh Miriam}[%
orcid=0000-0003-2523-6901,
email=miriam.doh@umons.ac.be,
]

\fnmark[1]
\address[1]{ISIA Lab - Université de Mons (UMONS)}
\address[2]{IRIDIA Lab - Université Libre de Bruxelles (ULB)}

\author[3]{Karagianni Anastasia}[%
orcid=0000-0003-4647-5311,
email=anastasia.karagianni@vub.be,
]
\fnmark[2]
\address[3]{Law, Science, Technology \& Society (LSTS) Research Group - Vrije Universiteit Brussels (VUB)}

\fntext[1]{This work was supported by the ARIAC project (No. 2010235), funded by the Service Public de Wallonie (SPW Recherche).}
\fntext[2]{This work was supported by the FARI - AI for the Common Good Institute (ULB-VUB), financed by the European Union, with the support of the Brussels
Capital Region (Innoviris and Paradigm)}

\begin{abstract}
This study delves into gender classification systems, shedding light on the interaction between social stereotypes and algorithmic determinations. Drawing on the "averageness theory," which suggests a relationship between a face's attractiveness and the human ability to ascertain its gender, we explore the potential propagation of human bias into artificial intelligence (AI) systems. Utilising the AI model Stable Diffusion 2.1, we have created a dataset containing various connotations of attractiveness to test whether the correlation between attractiveness and accuracy in gender classification observed in human cognition persists within AI. Our findings indicate that akin to human dynamics, AI systems exhibit variations in gender classification accuracy based on attractiveness, mirroring social prejudices and stereotypes in their algorithmic decisions. This discovery underscores the critical need to consider the impacts of human perceptions on data collection and highlights the necessity for a multidisciplinary and intersectional approach to AI development and AI data training. By incorporating cognitive psychology and feminist legal theory, we examine how data used for AI training can foster gender diversity and fairness under the scope of the AI Act\footnote{\url{https://www.europarl.europa.eu/news/en/press-room/20240308IPR19015/artificial-intelligence-act-meps-adopt-landmark-law.}} and GDPR\footnote{Regulation (EU) 2016/679 of the European Parliament and of the Council on the protection of natural persons with regard to the processing of personal data and on the free movement of such data, and repealing Directive 95/46/EC (General Data Protection Regulation). \url{https://eur-lex.europa.eu/legal-content/EN/TXT/PDF/?uri=CELEX:32016R0679}}, reaffirming how psychological and feminist legal theories can offer valuable insights for ensuring the protection of gender equality and non-discrimination in AI systems.

\end{abstract}


\maketitle

\section{Introduction}
\label{sec:introduction}
Language, the cornerstone of human interaction, encapsulates our thoughts, knowledge, experiences, and creative endeavours. It reflects our societies' historical and cultural complexities, perpetuating values, structures, and, inevitably, stereotypes \cite{1_de2019language,2_maass1999linguistic}. Stereotypes, first dissected in the Social Sciences by Lippman in 1922, delineate the “typical image” or representation conjured when referring to specific groups or situations \cite{3_bottom2012casual}. Despite their role in simplifying our understanding of the world, stereotypes often carry negative connotations, particularly in the realm of gender stereotypes, where gender identities are considered vastly different, echoing assumptions of biologically determined roles that dictate societal positions based on physical attributes or emotional capacities\cite{4_lefton2005psychology}.

Feminist theory critically examines gender stereotypes, exploring how social norms and expectations shape cis-normative perceptions of femininity and masculinity \cite{4_lefton2005psychology}. In particular, A. Oakley contributed to exploring gender as a social construct, challenging the reduction of gender to mere biological determinants and asserting its foundation in social, economic, and cultural constructs \cite{oakley1972sex}.

Given the deeply rooted nature of gender stereotypes in societal and cultural contexts, it is essential to examine how these biases are transferred into and expressed within digital technologies, particularly AI. While AI systems hold immense potential to revolutionize various aspects of our lives, they are not immune to embedding discrimination in subtle yet pervasive ways. Historically dominated by masculine perspectives, the technological field prompts concerns about inclusivity and the possibility that AI may perpetuate existing gender biases \cite{7_boivie2010women,8_bartoletti2022ai}.

For instance, in 2021, it was revealed that API image labelling services generated sexist labels [8]. In a dataset where all individuals had visible hair and wore professional attire, the Google Cloud Vision image labelling algorithm consistently paid more attention to women's hairstyles and fashion than men's, even though both had the same occupation as members of Congress.  These algorithms labelled men as "officials," "entrepreneurs," and "military officers," while women were often associated with labels like "hairstyle" or "beauty" \cite{schwemmer2020diagnosing}. Additionally, Microsoft's NSFW service detected female subjects as adult content at a higher rate \cite{MauroSchellmann2023}.

Such disparities underscore the urgency of our exploration into AI and gender bias, aiming to uncover whether these digital advancements are catering exclusively to male-dominated narratives or forging a path toward inclusivity.

Humans can introduce biases that become embedded in AI systems by determining which datasets, variables, and rules the algorithms learn from to make predictions. 
In this context, a critical approach towards the data collection process is sought to be adopted by our work, fully aware of how this can be influenced by human perception and acknowledged that according to a 2019 report by the European Union Agency for Fundamental Rights \cite{10_europaFundamentalRights}, data quality is an important risk factor for bias in AI.

As demonstrated in the aforementioned cases, the datasets used in these systems and their filtering processes highlight the interaction between words, images, and stereotypical connotations, which can negatively influence AI. To understand these mechanisms, this work aims to identify potential human biases in gender classification systems.

To lead this exploration, we adopt an approach aligned with the field of "Artificial Cognition" \cite{taylor2021artificial}. This area of research is based on the hypothesis that just as cognitive psychology has been used to understand the human mind as a 'black box', its theories can serve as a starting point for understanding the mechanisms of AI systems in their opacity. Consequently, this field suggests that by applying cognitive psychology theories to AI, we can gain insights into how human biases might be reflected in AI's decision-making processes.
Specifically, we explore the "averageness theory" \cite{langlois1990attractive,langlois1994average,rhodes2003fitting,rhodes1999average} within cognitive psychology, examining how perceptions of attractiveness could influence AI's gender classification accuracy. Through a dataset generated by the AI model Stable Diffusion 2.1, theoretical insight and empirical analysis are blended to scrutinise the manifestations of human gender bias in AI classification systems.

To contextualise our empirical investigation, we examine how the AI Act and GDPR address cognitive gender bias in classification systems. Moreover, this examination underscores the importance of regulation in addressing the legal challenges posed by generative AI-based data augmentation, particularly those related to gender bias and data protection principles such as collection limitation, purpose specification, use limitation, data minimisation, transparency, data quality, access and correction, retention limitation, automated decision-making, and profiling  \cite{11_cate2018artificial}.

This investigation, rooted in a deep understanding of gender dynamics based on the feminist legal theory \cite{12_levit2016feminist} and averageness theory, seeks to shed light on how unconscious assumptions and biases can shape future technologies, highlighting the need for careful ethical and legal consideration in the design and implementation of AI.

The work is structured as follows: in Section \ref{sec:introduction}, an introduction of the work is presented; in Section \ref{sec:Related Works}, state-of-the-art studies about bias in gender classification systems and text-to-image generation methods are presented; in Section \ref{sec:proposedidea} the core idea of the investigation is presented; in Section \ref{sec:ExpSetUp} the experiment set up is described; in Section \ref{sec:Experimental results}, the results of the experiments are included; in Section \ref{sec:FeministPers} provides a legal analysis of the issue; and in Section \ref{sec:conclusion}, the work is concluded.

\section{Related Works}
\label{sec:Related Works}

\subsection{Gender Bias in AI Classification Systems}

In transitioning to a detailed analysis of gender bias within AI systems, particularly in image gender classification, it's evident that these biases extend and amplify existing societal imbalances.
One of the critical works in the debate on gender bias in AI systems is "Gender Shade" \cite{buolamwini2018gender}, which revealed a trend of gender classification systems, such as those offered by APIs like Microsoft or IBM, showing more significant inaccuracies in classifications related to women and, in particular, women of colour. Furthermore, it emerged that such inaccuracies tend to increase with darker skin tones, highlighting a problematic combination of gender and racial biases. However, understanding the causes of these biases is a complex task. AI systems often operate as "black boxes," and the data used to train them are opaque, making it difficult to identify underlying behaviour.

Another consideration is that different gender classification services use different training data and infrastructure requirements. This suggests that each service may employ unique training data and have specific infrastructure requirements that influence gender classification \cite{scheuerman2019computers}. This diversity among services raises questions about the consistency and reliability of gender classifications produced by AI systems.
One of the main theories explaining gender bias in AI systems suggests that the discriminated demographic group may be underrepresented within the training data sets used for model training, as supported by the authors \cite{buolamwini2018gender}, showing that the most common data sets used in gender classification lacked diversity in terms of skin colour. However, this theory has been debunked by studies showing that balancing the training dataset did not eliminate bias \cite{wang2019balanced}. In the search for further causes, an important discovery is that skin type does not seem to be the determining factor in the accuracy of gender classification. The problem seems more complex and is related to the persistence of stereotypes within this classification. This could make the issue even more intersectional, as it could harm multiple social classes. For example, \cite{muthukumar2018understanding} highlighted that makeup and eye features are significant predictive factors for classifying a face as female, raising concerns about the perpetuation of gender stereotypes. Another study \cite{albiero2021gendered} has suggested that besides makeup, features such as hairstyle, facial structure, and clothing could be more relevant than skin type in determining gender, justifying that women were more prone to false non-match rate (FNMR) than men in a face recognition context. \cite{grother2019face} reported that gender differences in FNMR are not universal. Gender social conventions related to hairstyle and makeup, by definition, can vary significantly among social groups, so it seems likely that they manifest in various ways. Social conventions related to hairstyle and makeup also change with a person's age, and therefore, they play a role in understanding how facial recognition accuracy varies across age groups.

Furthermore, an interesting study \cite{scheuerman2019computers} attempted to analyse these algorithms by considering the transgender community and trying to classify transwomen and transmen. One of the most exciting results of this research is that transgender men had the lowest true positives in gender classification, suggesting that there is a more \textit{cis-normative} male representation for the male category, which is less subject to variety and diversity in the training data. Outside the world of research and computer science, the public has also begun to wonder about the causes of misclassifications or certain system decisions. In 2021, an algorithmic artist and transgender individual, Ada Ada Ada \cite{ada-ada-ada}, attempted to test how algorithms perceive gender. By using her own transgender body, the artist found several methods for tricking
gender recognition technology into seeing a gender different from the initial judgment. For example, she discovered how varying her emotional expression, head tilt, hair and beard, eyes, and nose could lead to being classified as a specific gender. This result comes after a series of machine tests and post-analysis of the results obtained. This project once again demonstrates how these classifications are based on social stereotypes. 

\subsection{BIAS in image generation}
Regarding text-to-image (TTI) systems, academic inquiry has underscored a marked prevalence of demographic biases, especially those pertaining to gender and race. These biases are evidenced through stereotypical depictions across assorted domains, such as vocations and personal traits, underscoring an inclination towards the over-representation of attributes linked with whiteness and masculinity \cite{25_luccioni2023stable,27_naik2023social}. 
Examinations have elucidated that a principal factor contributing to these biases is the instructional content utilised to train models, typically sourced from the internet, which mirrors the stereotypes and prejudices extant within society \cite{25_luccioni2023stable}. Despite recognising their inherent biases, employing models like CLIP \cite{agarwal2021evaluating} to steer the generative process in apparatuses such as Stable Diffusion further exacerbates the issue, as it amplifies the perpetuation of biases \cite{49__radford2021learning}. Furthermore, the exploration of representations engendered by TTI models has divulged that biases and stereotypes are not confined to the portrayal of individuals but also extend to objects, clothes, and even national identities \cite{wu2023stable}, reflecting a wide spectrum of demographic biases. 
Whilst endeavours have been undertaken to ameliorate these biases, for instance, through the analysis of models' latent spaces to render the generated images more representative, the efficacy of such measures remains in question \cite{14__brack2022stable,55__schramowski2023safe}.

\section{Proposed idea: From Human Bias to Machine Bias}
\label{sec:proposedidea}
This work is inspired by the theories of cognitive psychology concerning gender perception and embarks upon a translational exploration from human to machine. 

Gender discrimination within human cognitive processes has been extensively probed by cognitive psychology, with initial examinations focusing on how distinct facial features between males and females influence gender perception. Studies such as \cite{sanderson_1975,nakdimen1984physiognomic} have illuminated that facial dimensions, nose shape, prominence of jaws and eyebrows, and the structure of cheekbones contribute to gender perception, yet it is acknowledged that no single trait definitively dictates it. Indeed, it is highlighted that the complexity of gender perception surpasses mere physiognomy, showcasing how generalisations can readily evolve into stereotypes. This nuance in defining gender underscores the necessity of critically evaluating stereotypical visual representations of men and women—a phenomenon that has been extensively documented across various media \cite{kerkhoven2016gender,amini2012gender,jang2019quantification}, yet remains underexplored in the digital domain \cite{singh2020female}.

In this context, a pivotal role is played by the averageness theory, suggesting that faces deemed attractive, due to their prototypical nature, are more easily classified by gender \cite{langlois1990attractive,langlois1994average,rhodes2003fitting,rhodes1999average}. This theory is supported by findings that facial attractiveness facilitates classification in adults. However, this correlation is observed to vary across genders, with attractive female faces generally perceived as highly feminine, unlike their male counterparts \cite{o1998perception,rhodes2000sex,langlois2000maxims}. These dynamics introduce the concept of "face space", where faces are categorised in a multidimensional space based on the variance of facial traits and their encoding \cite{valentine1991unified}, with attractive and prototypically feminine faces positioned at the centre of this space.

Building on these foundations, our study explores how cognitive biases manifest within AI systems, specifically focusing on gender classification mechanisms. We delve into the potential propagation of gender stereotypes and investigate the role of attractiveness in classification accuracy. Recent research has largely focused on identifying biases related to gender, ethnicity, makeup usage, and skin colour. However, our approach aims to innovate by emphasising the impact of attractiveness—a composite attribute influenced by multiple facial features—which plays a crucial role in the perception and classification of gender.

Our analysis of bias is approached from two dimensions:
\begin{itemize}
    \item \textit{First Level of Representation} \cite{noble2018}: We question whether classification systems consistently achieve the same level of accuracy across all analysed groups (attractive/unattractive, women/men).
    \item \textit{Conditional Demographic Parity} \cite{corbett-davies2017}: We consider scenarios where biases may arise if the system systematically produces only a subset of possible labels, even if the algorithm's output is correct. For example, if men and women in a sample are dressed similarly, an unbiased algorithm would be expected to return the "clothing" label with equal frequency for each gender.
\end{itemize}
To summarise, our primary research question is:

\textbf{RQ1:} How does the averageness theory influence the performance of AI algorithms in gender classification?

Several secondary questions support this:
\begin{itemize}
    \item \textbf{RQ1.a:} Is there a difference in classification accuracy between attractive and unattractive faces within gender groups in AI algorithms?
    \item \textbf{RQ1.b:} Do the performances of AI algorithms in gender classification maintain uniformity across different demographic groups, particularly when considering attractiveness?
    \item \textbf{RQ1.c:} Do gender classification algorithms exhibit gender stereotypes, and in what forms do these stereotypes manifest?

\end{itemize}

\section{Experimental Setup}
\label{sec:ExpSetUp}
\subsection{Rationale for Synthetic Dataset Generation}
When addressing the subject of attraction, a frequent critique is raised concerning its inherently subjective nature. However, it is noteworthy that datasets have been developed to tackle this aspect, exemplified by the HotOrNot dataset \cite{white2004automatic,gray2010predicting,donahue2011annotator}, the SCUT-FBP dataset \cite{xie2015SCUT_FBP}, and the CelebA dataset \cite{zhang2020celeba}.
In particular, The HotOrNot dataset was created by collecting user ratings of attractiveness from the HotOrNot website, a site launched in the 2000s where users rated pictures of individuals on a scale from 1 to 10, generating a large dataset of images paired with attractiveness scores. For example, a dataset version was formally presented in \cite{donahue2011annotator}, where researchers used it to improve image annotation in attractiveness task scenarios.
The SCUT-FBP dataset was developed to provide a more systematic set of facial images for beauty prediction tasks. Like the previous dataset, SCUT-FBP contains facial images with beauty scores annotated by multiple human raters. This dataset aimed to create a more controlled environment for studying facial attractiveness by ensuring in-scene and face expression conditions. Lastly, the CelebA dataset is a large-scale face attributes dataset with celebrity images, each annotated with 40 attribute labels, including one related to attractiveness. Unlike the previous two datasets, CelebA was designed to facilitate general research in face analysis (detection, attribute prediction, and other fields like beauty prediction).

Despite these efforts, datasets in this domain exhibit significant variability, especially a lack of representation across various ethnicities or groups of people (e.g., SCUT-FBP is limited to 500 samples and focuses primarily on Asian and Caucasian faces, or CelebA contains only celebrity individuals). Moreover, HotOrNot is based on data collected from a website where people voluntarily uploaded photos of themselves to be rated; there was no control over the data type collected. This resulted in a lack of consistent distribution across ethnicity and gender, leading to an uncontrolled environment.

\begin{figure}[h]
\centering
\includegraphics[width=0.8\textwidth]{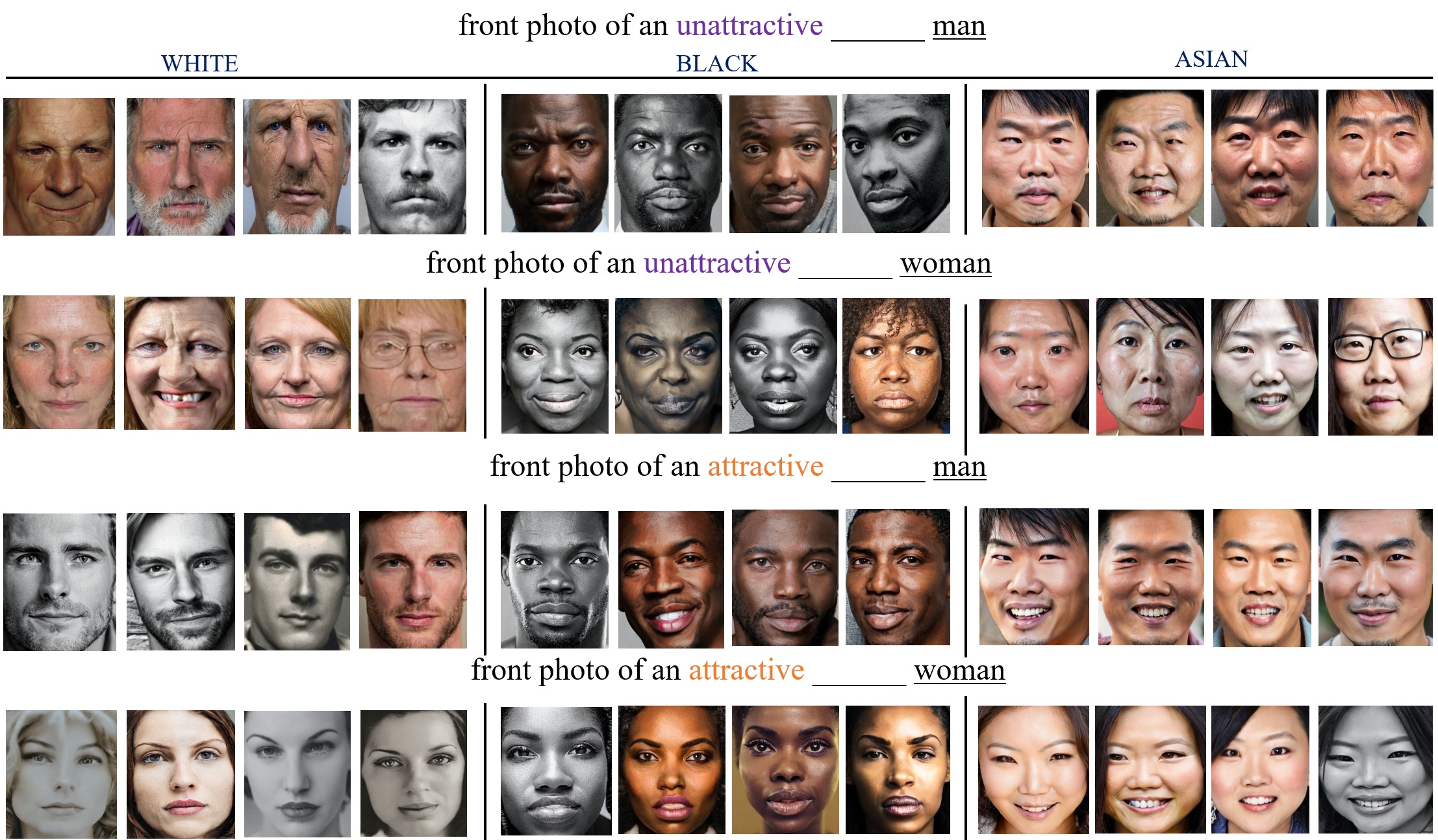}
\caption{Examples of sample images in the created dataset using Stable Diffusion 2.1 for the prompts 'front photograph of an unattractive/attractive \textit{ethnicity} man/woman.' for White, Black, and Asian groups}
\label{fig:aw}
\end{figure}

Consequently, for this study, a compact synthetic dataset has been compiled in light of these limitations, generated using Stable Diffusion \cite{rombach2022highresolution, StableDiffusionHuggingFace} as shown in Figure \ref{fig:aw}.

This approach's choice is deeply rooted in Stable Diffusion’s comprehensive training across diverse image and label datasets. This training equips the model with a grasp of the subjective nuances of human attractiveness, allowing it to reflect the varied interpretations of facial attractiveness found in datasets. Moreover, the rationale for using a synthetic dataset lies in the ability to control and vary the generated images' attributes systematically. This approach addresses the limitations of existing datasets by ensuring a balanced representation of different ethnicities and providing a consistent framework for studying the subjective aspects of attractiveness. Additionally, synthetic data is gaining traction in data augmentation \cite{chen2023unified,}, making it intriguing to explore the type of representations these datasets can provide for this study.
Finally, this method allows the investigation of the two dimensions of bias presented in section \ref{sec:proposedidea}. In particular, these dimensions are analysed by testing various gender classification models, observing the variation in accuracy (\textit{First Level of Representation}), and qualitatively analysing the results generated by Stable Diffusion for the requested prompt (\textit{Conditional Demographic Parity}).
\subsection{Dataset Creation and Processing}
As mentioned, to create a balanced and diverse dataset, Stable Diffusion was tasked with generating images based on the following prompts: \textit{'frontal photograph of an attractive/unattractive \textit{ethnicity} man/woman'}. Within these prompts, the descriptor \textit{ethnicity} was systematically alternated with "White", "Black", and "Asian", ensuring a diverse representation of ethnicities within the resulting dataset.

Following this approach, the Stable Diffusion API was utilised to generate the dataset; specifically, version 2.1 of the stability/stable-diffusion model \cite{StableDiffusionHuggingFace}, accessible through the demo provided by Hugging Face, was employed. The default guidance scale value of 9 was kept. This process resulted in 200 images per class, resulting in 2400 images.
\begin{table}[h]
\centering
\caption{Representation of class percentage in the dataset created.}
\begin{tabular}{lcc}
\hline
\textbf{Ethnicity} & \textbf{Attractive} & \textbf{Unattractive} \\ \hline
Asian              & 16.59\%             & 16.62\%               \\
Black              & 16.16\%             & 16.54\%               \\
White              & 17.64\%             & 16.11\%               \\ \hline
\label{tab:rep}
\end{tabular}
\end{table}

The images were subsequently cropped to focus on the face, using the Multi-Task
Cascade CNN (MTCNN) \cite{xiang2017joint} face detection algorithm to standardise the size of the face portions in all images. This process made lose some of the sample generated since the face was not detect during the process. The final dataset contains 2324 crop images distributed according to specific proportions, as shown in Table \ref{tab:rep}. 
\subsection{Gender classification models and Metrics}
Regarding the accuracy analysis, the models considered are those offered by Amazon Rekognition by Amazon Web Services (AWS) \cite{amazonrekognition}, the DeepFace library \cite{deepfaceRepo}, and the InsightFace one \cite{insightfaceRepo}. Our selection included a commercial API, Amazon, and two projects widely recognised on GitHub: DeepFace (with 9.6k stars) and InsightFace (with 20.9k stars). Among these, DeepFace is the only one that provides data on the accuracy of its gender recognition model, achieving an accuracy of 97.44\%.

To evaluate the model performances we take into account the following metrics:
\begin{itemize}
    \item PPV (Positive Predictive Value): Positive Predictive Value, or PPV, is a metric used to analyse diagnostic test results or predictive models. In this formula, "TP" stands for true positives, cases where the test or model correctly predicted membership in a category. In contrast, "FN" stands for false negatives, where the test or model incorrectly predicted that an item does not belong to a class when it does. PPV measures the proportion of correct positive predictions relative to the total positive predictions and is expressed as a percentage.
    \begin{align}
    PPV &= \frac{TP}{TP + FN}  
    \end{align}
    \item ER (Error Rate): Error Rate is another metric used to evaluate the performance of tests or prediction models. In this formula, "FP" stands for false positives, i.e., cases where the test or model erroneously predicted membership in a category when it does not, and "FN" still represents false negatives. Error Rate measures the proportion of incorrect predictions relative to the total predictions and is expressed as a percentage. A lower error rate indicates a higher accuracy of the test or model.
    \begin{align}
    Error\ Rate &= \frac{FP + FN}{TP + FP + FN} 
    \end{align}
\end{itemize}

\section{Experimental results}
\label{sec:Experimental results}
\subsection{Analysis of Gender Classification Based on Accuracy - \textit{"First level of representation"} (RQ1.a. and RQ1.b)}
\begin{table}[htbp]
    \centering
    \caption{Gender classification performance for Amazon Rekognition, measured in terms of Positive Predictive Value (PPV) and error rate.}
    \label{tab:amazonrekognition}
    \begin{tabular}{|l|l|c|c|c|c|}
        \hline
         & & \multicolumn{4}{c|}{Amazon Rekognition} \\
        \hline
        Group& Metric (\%) & \textbf{Asian} & \textbf{Black} & \textbf{White} & \textbf{Avg.} \\
        \hline
        \multirow{2}{*}{\textbf{A-M}} & PPV & 100 & 99.46 & 95.07 & \textbf{98.18} \\
        \cline{2-6}
        & ER & \cellcolor{red!0}- & \cellcolor{red!0.53}0.53 & \cellcolor{red!5}4.93 & \cellcolor{red!2} \textbf{1.82} \\
        \hline
        \multirow{2}{*}{\textbf{U-M}} & PPV & 99.45 & 100 & 99.00 & \textbf{99.48} \\
        \cline{2-6}
        & ER & \cellcolor{red!0.54}0.54 & \cellcolor{red!0}- & \cellcolor{red!1}1.00 & \cellcolor{red!0}\textbf{0.52} \\
        \hline
        \multirow{2}{*}{\textbf{A-W}} & PPV & 100 & 100 & 100 & \textbf{100} \\
        \cline{2-6}
        & ER & \cellcolor{red!0}- & \cellcolor{red!0}- & \cellcolor{red!0}- & \cellcolor{red!0}\textbf{-} \\
        \hline
        \multirow{2}{*}{\textbf{U-W}} & PPV & 75.77 & 99.49 & 91.63 & \textbf{88.96} \\
        \cline{2-6}
        & ER & \cellcolor{red!24.23}24.23 & \cellcolor{red!0.51}0.51 & \cellcolor{red!8.37}8.37 & \cellcolor{red!10}\textbf{11.04} \\
        \hline
    \end{tabular}

\end{table}
The metrics of interest spanned gender, attractiveness, and ethnicity, as shown in the tab. \ref{tab:amazonrekognition}, \ref{tab:combined_horizontal_separated} where A stands for attractive, "U" for unattractive, "M" for men and "W" for women. 
An analysis of gender classification revealed a noticeable difference in model performance, particularly when examining the gradient of attractiveness.
Whether deemed attractive or not, the model's accuracy showed little variation for male subjects. However, when classifying female subjects, the models displayed a change in accuracy between attractive and unattractive groups. InsightFace's Positive Predictive Value (PPV) decreased from 85.61\% for attractive women to 62.74\% for those considered unattractive. The disparity was even more pronounced for DeepFace, which saw a PPV drop from 67.5\% for attractive women to 21.22\% for unattractive women. Amazon Rekognition generally emerged as the most robust model; nonetheless, a slight performance difference was observed: from 100\% PPV for attractive women to 88.96\% for unattractive women. Regarding ethnicity, the most disadvantaged group for InsightFace and DeepFace was unattractive black women, with an error rate of 44.44\% for InsightFace and 85.86\% for DeepFace. Conversely, for Amazon Rekognition, black subjects were less disadvantaged, while unattractive Asian women showed a more significant performance degradation.

While InsightFace and DeepFace showed variable performance between attractive and unattractive females, Amazon Rekognition consistently maintained high accuracy across gender and attractiveness attributes.

\begin{table}[htbp]
    \centering
    \caption{Comparison of gender classification performance between InsightFace and DeepFace, measured in terms of Positive Predictive Value (PPV) and Error Rate (ER).}
    \label{tab:combined_horizontal_separated}
    \begin{tabular}{|l|l|c|c|c|c||c|c|c|c|}
        \hline
         & & \multicolumn{4}{c||}{InsightFace} & \multicolumn{4}{c|}{DeepFace} \\
        \hline
         Group & Metric (\%)  & \textbf{Asian} & \textbf{Black} & \textbf{White} & \textbf{Avg.} & \textbf{Asian} & \textbf{Black} & \textbf{White} & \textbf{Avg.} \\
        \hline
        \multirow{2}{*}{A-M} & PPV & 79.45 & 93.54 & 91.00 & \textbf{87.98} & 98.95 & 99.46 & 100 &  \textbf{99.47} \\
        & ER & \cellcolor{red!20.54}20.54 & \cellcolor{red!6.45}6.45 & \cellcolor{red!9}9.00 & \cellcolor{red!11.99}\textbf{11.99} & \cellcolor{red!1.04}1.04 & \cellcolor{red!0.53}0.53 & \cellcolor{red!0}- & \cellcolor{red!0}\textbf{0.52} \\
        \hline
        \multirow{2}{*}{U-M} & PPV & 85.86 & 91.39 & 87.19 & \textbf{88.14} & 100 & 100 & 100 &  \textbf{100} \\
        & ER & \cellcolor{red!14.13}14.13 & \cellcolor{red!8.60}8.60 & \cellcolor{red!12.80}12.80 & \cellcolor{red!12}\textbf{11.84} & \cellcolor{red!0}- & \cellcolor{red!0}- & \cellcolor{red!0}- & \cellcolor{red!0}\textbf{-} \\
        \hline
        \multirow{2}{*}{A-W} & PPV & 96.41 & 75.77 & 84.65 & \textbf{85.61 }& 72.82 & 53.09 & 76.72 &  \textbf{67.54} \\
        & ER & \cellcolor{red!3.58}3.58 & \cellcolor{red!24.22}24.22 & \cellcolor{red!15.34}15.34 & \cellcolor{red!14.38}\textbf{14.38} & \cellcolor{red!27.18}27.18 & \cellcolor{red!46.91}46.91 & \cellcolor{red!23.28}23.28 & \cellcolor{red!32.46}\textbf{32.46} \\
        \hline
        \multirow{2}{*}{U-W} & PPV & 70.61 & 55.55 & 62.06 & \textbf{62.74} & 17.01 & 14.14 & 32.51 &  \textbf{21.22} \\
        & ER & \cellcolor{red!29.38}29.38 & \cellcolor{red!44.44}44.44 & \cellcolor{red!37.93}37.93 & \cellcolor{red!37.25}\textbf{37.25} & \cellcolor{red!82.99}82.99 & \cellcolor{red!85.86}85.86 & \cellcolor{red!67.49}67.49 & \cellcolor{red!78.78}\textbf{78.78} \\
        \hline
    \end{tabular}
\end{table}

\begin{figure}[h]
    \centering
    \includegraphics[width=0.8\textwidth]{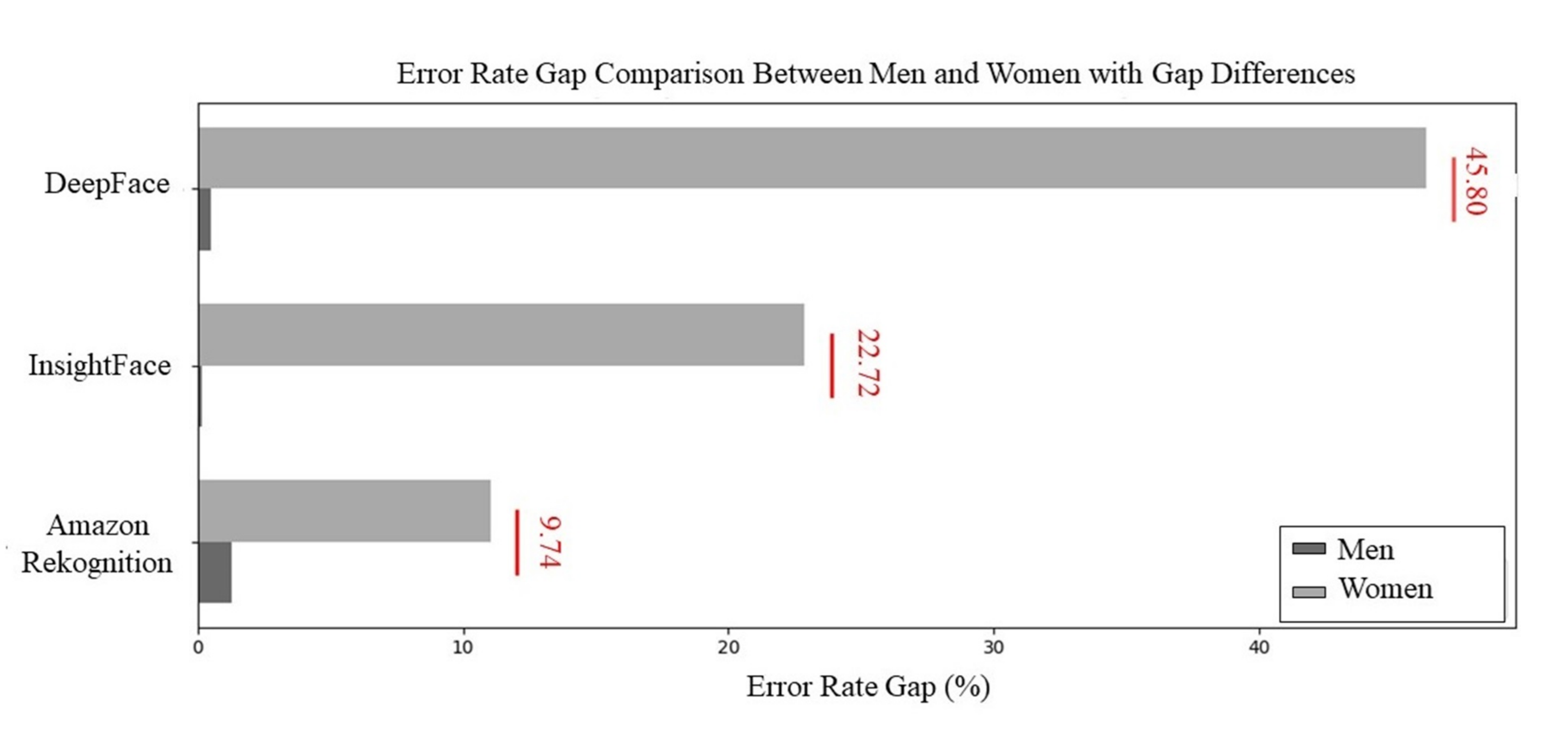}
    \caption{Error Rate gap between Attractive and Unattractive Men/Women for the models analysed. The red line reports the inner gap between the female and male error gap for each model.}
    \label{fig:gender gap}
\end{figure}

The calculated error rate disparities across Amazon Rekognition, InsightFace, and DeepFace models underscore distinct biases concerning gender and perceived attractiveness, as documented in the figure \ref{fig:gender gap}.
InsightFace exhibited a minimal gap for men, suggesting uniform performance across levels of attractiveness. However, a substantial disparity was observed for women, with unattractive women experiencing significantly higher error rates. DeepFace revealed the starkest contrast in error rates among women, with unattractive women facing dramatically higher error rates, highlighting a potential bias towards attractiveness in female gender recognition. Deepface presents a higher difference between the error rate gaps between the male and female samples, 45.80.

These findings indicate a persistent trend where women, especially those categorised as unattractive, are at a disadvantage due to higher error rates.

\subsection{Quantitative Analysis of Physical Characteristics and Facial Expressions in a Stable Diffusion-Generated Face Dataset - \textit{"Conditional demographic parity"} (RQ1.c)}

To understand the kind of physiognomy generated by the Stable Diffusion model in response to the prompt, a qualitative analysis of the dataset was conducted by examining images of individuals, leading to several observations.

\begin{figure}[h]
    \centering
    \includegraphics[width=0.8\textwidth]{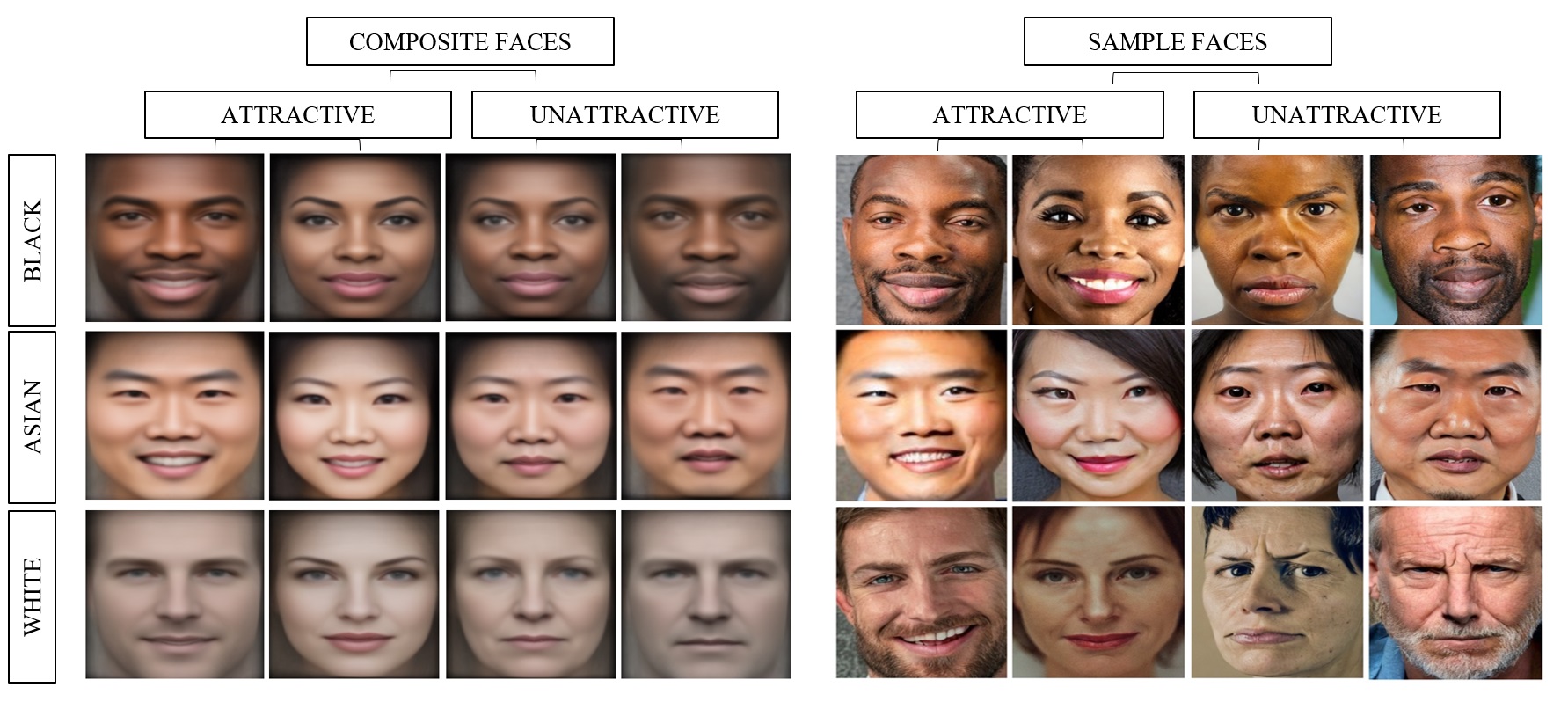}
    \caption{Composite faces and face samples of attractive/unattractive men/women with clear differences in makeup use, face expression and age gap.}
    \label{fig:af}
\end{figure}

Initially, average faces were created by overlaying images of various groups of attractive and non-attractive individuals, both women and men. This analysis observed expressive differences between averagely attractive and non-attractive faces. The average attractive face invariably appears smiling or calm, whereas the average non-attractive face tends to exhibit a more severe expression. Furthermore, attractive average faces seem younger than their non-attractive counterparts (fig. \ref{fig:af}).
From this initial analysis of averages, with scrutiny applied to each image in turn, three interesting observations have emerged:
\begin{itemize}
    
    \item \textbf{Differences in makeup:} There is a significant incidence of makeup on the average attractive and non-attractive face of women. Attractive women show signs of makeup with a more pronounced application, especially on the lips. On the other hand, non-attractive women have lighter or even absent makeup, especially Asian women. An exception is Black women, who are rarely generated without makeup regardless of attractiveness.

  \item \textbf{Similarities among Black subjects:}It is interesting to note that there are no significant differences between the average attractive and non-attractive faces for Black men and women, except for a slightly more smiling expression in attractive Black subjects. Moreover, Black subjects are never generated without a beard, as shown by the average faces, which are attractive and non-attractive in the presence of the beard.

  \item \textbf{Age Gap:} Among women of all ethnicities, youth appears to be a distinctive trait of attractiveness. White women deemed attractive tend to show youthful visual characteristics, with rare exceptions of white hair, suggesting a strong link between youthfulness and the perception of beauty. Conversely, non-attractive women of the same ethnicity appear older, with a greater frequency of white hair, suggesting that ageing may negatively impact their aesthetic perception. A similar trend is observed among Asian women, where youthful features prevail among those considered attractive, while their non-attractive counterparts show more marked signs of ageing. However, Black women seem to follow the trend but with a less pronounced age gap. Regarding men, the link between age and attractiveness manifests less markedly than in women but remains significant. Attractive white men can vary in age, displaying both youthful characteristics and signs of ageing, such as white hair or wrinkles, indicating a broader range of attractive traits. Non-attractive men, however, tend to be characterised by a seemingly more advanced age. This pattern of age-related variation is also reflected among Asian men. As with women, the trend is respected for Black men but with a smaller gap.

\end{itemize}

After conducting the primary analysis, it was decided to validate the observations concerning makeup differences and the age gap using attribute classification systems. Amazon Rekognition was utilized for the age attribute, as the API also provided age detection. Among all the evaluated models, it exhibited the fewest errors in gender classification. A trained lightened moon Mxnet model was employed for facial attributes \cite{mxnetFaceRepo}, especially the pre-trained on the CelebA dataset's attribute labels \cite{liu2015faceattributes}.

\begin{itemize}
    \item \textbf{Age analysis:} Amazon Rekognition allows the detection of an age class. Since this range is variable, to facilitate the analysis, we decided to adopt a set of fixed ages ranging from 0 to "$\geq70$", proceeding by decades as reported in the tab.
\ref{tab:consolidated_distribution}.
 With this age range set, a clear trend emerges regarding age between attractive and non-attractive women, as observed visually from a qualitative analysis. Non-attractive women tend to distribute in older age bands, with a significant presence among those aged 40-49 years (36.95\%) and 50-59 years (37.93\%) and some samples over 70 years. In contrast, attractive women are generally younger, with about 60\% of the samples in the 20-29 year age range. This trend is consistent across all ethnicities. Furthermore, it is possible to see that generally, Asian women are depicted as younger compared to all other groups. At the same time, non-attractive women are older than the others. A similar pattern is observed for men, with non-attractive men slightly older than the attractive ones. The most attractive men belong to the age range of 20 to 39 years, while non-attractive men are predominantly in the age range of 30 to 59 years. Ethnic differences follow a similar trend, with slight variations in age range distributions. Once again, Asians are seen as younger by the model and white men as older. A strong connection between attractiveness and age is evident, with attractive individuals appearing younger than non-attractive ones, particularly among women.
    \begin{table}[htbp]
\centering
\caption{Consolidated Distribution of Samples Based on Age Ranges for Attractive (A) and Unattractive (U) Men (M) and Women (W) Across White, Black, and Asian ethnicities by AmazonRekognition}
\label{tab:consolidated_distribution}
\begin{tabular}{|l|cc|cc|cc|cc|cc|cc|}
\hline
\rowcolor{gray!50}
\multicolumn{1}{|c|}{\textbf{Age}} & \multicolumn{6}{c|}{\textbf{Men}} & \multicolumn{6}{c|}{\textbf{Women}} \\ 
\rowcolor{gray!25}
\multicolumn{1}{|c|}{} & \multicolumn{2}{c|}{\textbf{White}} & \multicolumn{2}{c|}{\textbf{Black}} & \multicolumn{2}{c|}{\textbf{Asian}} & \multicolumn{2}{c|}{\textbf{White}} & \multicolumn{2}{c|}{\textbf{Black}} & \multicolumn{2}{c|}{\textbf{Asian}} \\ 
\multicolumn{1}{|c|}{} & \textbf{A (\%)} & \textbf{U (\%)} & \textbf{A (\%)} & \textbf{U (\%)} & \textbf{A (\%)} & \textbf{U (\%)} & \textbf{A (\%)} & \textbf{U (\%)} & \textbf{A (\%)} & \textbf{U (\%)} & \textbf{A (\%)} & \textbf{U (\%)} \\ \hline
$\geq$70 & & & & & & & & \cellcolor{pink!5}0.49 & & & & \\
60-69 & & \cellcolor{blue!20}2.00 & & & & \cellcolor{blue!5}0.54 & & \cellcolor{pink!20}1.97 & & & & \\
50-59 & & \cellcolor{blue!32}32.00 & & & & \cellcolor{blue!43}4.32 & & \cellcolor{pink!38}37.93 & & \cellcolor{pink!10}1.01 & & \cellcolor{pink!10}1.03 \\
40-49 & \cellcolor{blue!18}17.54 & \cellcolor{blue!36}36.50 & \cellcolor{blue!3}2.69 & \cellcolor{blue!27}26.88 & \cellcolor{blue!1}1.05 & \cellcolor{blue!53}52.97 & & \cellcolor{pink!37}36.95 & \cellcolor{pink!2}1.55 & \cellcolor{pink!39}38.89 & \cellcolor{pink!5}0.51 & \cellcolor{pink!19}18.56 \\
30-39 & \cellcolor{blue!58}58.13 & \cellcolor{blue!27}26.50 & \cellcolor{blue!63}62.90 & \cellcolor{blue!60}60.22 & \cellcolor{blue!43}42.93 & \cellcolor{blue!31}30.81 & \cellcolor{pink!26}26.46 & \cellcolor{pink!13}13.30 & \cellcolor{pink!37}37.11 & \cellcolor{pink!47}46.97 & \cellcolor{pink!6}5.64 & \cellcolor{pink!47}47.42 \\
20-29 & \cellcolor{blue!24}24.14 & \cellcolor{blue!4}4.00 & \cellcolor{blue!34}34.41 & \cellcolor{blue!13}12.90 & \cellcolor{blue!37}36.65 & \cellcolor{blue!8}7.57 & \cellcolor{pink!68}67.72 & \cellcolor{pink!9}9.36 & \cellcolor{pink!61}60.82 & \cellcolor{pink!13}13.13 & \cellcolor{pink!53}52.82 & \cellcolor{pink!24}24.23 \\
20-19 & \cellcolor{blue!1}0.49 & & & & \cellcolor{blue!19}19.37 & \cellcolor{blue!4}3.78 & \cellcolor{pink!2}2.12 & & \cellcolor{pink!1}0.52 & & \cellcolor{pink!41}41.03 & \cellcolor{pink!9}8.76 \\
0-9 & & & & & & & & & & & & \\ \hline
\end{tabular}
\end{table}

    \item \textbf{Attribute analysis:} Some earlier observations can be validated by examining the datasets through attribute detection. Initially, the age gap is validated by the observation that for both women and men, the percentage of the "Young" attribute decreases when moving from attractive to unattractive subjects, supporting the notion that unattractive individuals are typically older. Furthermore, the difference in makeup application is also evident. Among attractive women, "Wearing Lipstick" and "Heavy Makeup" are observed as among the top four detected attributes, with respective percentages of 91.50\% and 71.74\%. Conversely, for unattractive women, makeup does not present as a significantly detected attribute, with only a "Wearing Lipstick" percentage of 16.15\% being observed (tab \ref{tab:att_distribution}).

\end{itemize}
\begin{table}[]
\caption{Average top-10 attributes detected for Attractive/Unattractive (A/U) Women (W) and Attractive/Unattractive Men (M). The attributes are generalised into 5 main groups: Hair Attributes (purple), Makeup and Accessories (orange), Facial Expression and Features (light blue), Beard Attributes (green) and Other Physical Attributes (yellow).}
\label{tab:att_distribution}
\begin{tabular}{|cc|cc|cc|cc|}
\hline
\multicolumn{2}{|c|}{A-M}                                                & \multicolumn{2}{c|}{U-M}                                              & \multicolumn{2}{c|}{A-W}                                               & \multicolumn{2}{c|}{U-W}                                               \\ \hline
\multicolumn{1}{|c|}{Attributes}                                 & (\%)  & \multicolumn{1}{c|}{Attributes}                               & (\%)  & \multicolumn{1}{c|}{Attributes}                                & (\%)  & \multicolumn{1}{c|}{Attributes}                                & (\%)  \\ \hline
\multicolumn{1}{|c|}{\cellcolor[HTML]{96FFFB}Big\_Lips}          & 96,77 & \multicolumn{1}{c|}{\cellcolor[HTML]{96FFFB}Big\_Lips}        & 96,24 & \multicolumn{1}{c|}{\cellcolor[HTML]{FFFFC7}Young}             & 100   & \multicolumn{1}{c|}{\cellcolor[HTML]{9AFF99}No\_Beard}         & 95,79 \\ \hline
\multicolumn{1}{|c|}{\cellcolor[HTML]{FFFFC7}Young}              & 95,86 & \multicolumn{1}{c|}{\cellcolor[HTML]{96FFFB}Big\_Nose}        & 64,00 & \multicolumn{1}{c|}{\cellcolor[HTML]{9AFF99}No\_Beard}         & 98,80 & \multicolumn{1}{c|}{\cellcolor[HTML]{FFFFC7}Young}             & 77,78 \\ \hline
\multicolumn{1}{|c|}{\cellcolor[HTML]{96FFFB}High\_Cheekbones}   & 73.08 & \multicolumn{1}{c|}{\cellcolor[HTML]{FFFFC7}Young}            & 57,42 & \multicolumn{1}{c|}{\cellcolor[HTML]{FFCE93}Wearing\_Lipstick} & 91,50 & \multicolumn{1}{c|}{\cellcolor[HTML]{96FFFB}High\_Cheekbones}  & 65,24 \\ \hline
\multicolumn{1}{|c|}{\cellcolor[HTML]{96FFFB}Smiling}            & 77,62 & \multicolumn{1}{c|}{\cellcolor[HTML]{9AFF99}No\_Beard}        & 54,51 & \multicolumn{1}{c|}{\cellcolor[HTML]{FFCE93}Heavy\_Makeup}     & 71,74 & \multicolumn{1}{c|}{\cellcolor[HTML]{CBCEFB}Black\_Hair}       & 49,42 \\ \hline
\multicolumn{1}{|c|}{\cellcolor[HTML]{CBCEFB}Black\_Hair}        & 54,63 & \multicolumn{1}{c|}{\cellcolor[HTML]{96FFFB}High\_Cheekbones} & 50,47 & \multicolumn{1}{c|}{\cellcolor[HTML]{96FFFB}High\_Cheekbones}  & 63,64 & \multicolumn{1}{c|}{\cellcolor[HTML]{96FFFB}Big\_Lips}         & 48,23 \\ \hline
\multicolumn{1}{|c|}{\cellcolor[HTML]{96FFFB}Big\_Nose}          & 51,54 & \multicolumn{1}{c|}{\cellcolor[HTML]{CBCEFB}Black\_Hair}      & 46,72 & \multicolumn{1}{c|}{\cellcolor[HTML]{96FFFB}Smiling}           & 57,56 & \multicolumn{1}{c|}{\cellcolor[HTML]{96FFFB}Smiling}           & 44,91 \\ \hline
\multicolumn{1}{|c|}{\cellcolor[HTML]{9AFF99}No\_Beard}          & 46,39 & \multicolumn{1}{c|}{\cellcolor[HTML]{9AFF99}Goatee}           & 38,20 & \multicolumn{1}{c|}{\cellcolor[HTML]{CBCEFB}Black\_Hair}       & 53,54 & \multicolumn{1}{c|}{\cellcolor[HTML]{96FFFB}Big\_Nose}         & 41,09 \\ \hline
\multicolumn{1}{|c|}{\cellcolor[HTML]{9AFF99}5\_o\_Clock Shadow} & 44,83 & \multicolumn{1}{c|}{\cellcolor[HTML]{96FFFB}Smiling}          & 33,53 & \multicolumn{1}{c|}{\cellcolor[HTML]{96FFFB}Big\_Nose}         & 40,47 & \multicolumn{1}{c|}{\cellcolor[HTML]{FFCE93}Wearing\_Lipstick} & 16,15 \\ \hline
\multicolumn{1}{|c|}{\cellcolor[HTML]{9AFF99}Goatee}             & 30,49 & \multicolumn{1}{c|}{\cellcolor[HTML]{9AFF99}Mustache}         & 22,58 & \multicolumn{1}{c|}{\cellcolor[HTML]{96FFFB}Big\_Lips}         & 34,41 & \multicolumn{1}{c|}{\cellcolor[HTML]{CBCEFB}Bangs}             & 10,49 \\ \hline
\multicolumn{1}{|c|}{\cellcolor[HTML]{CBCEFB}Bushy\_Eyebrows}    & 28,38 & \multicolumn{1}{c|}{\cellcolor[HTML]{FFFFC7}Chubby}           & 14,08 & \multicolumn{1}{c|}{\cellcolor[HTML]{CBCEFB}Wavy\_Hair}        & 21,09 & \multicolumn{1}{c|}{\cellcolor[HTML]{CBCEFB}Brown\_Hair}       & 9,89  \\ \hline
\end{tabular}
\end{table}
\subsection{Controversial images}
\begin{figure}[h]
    \centering
    \includegraphics[width=0.6\textwidth]{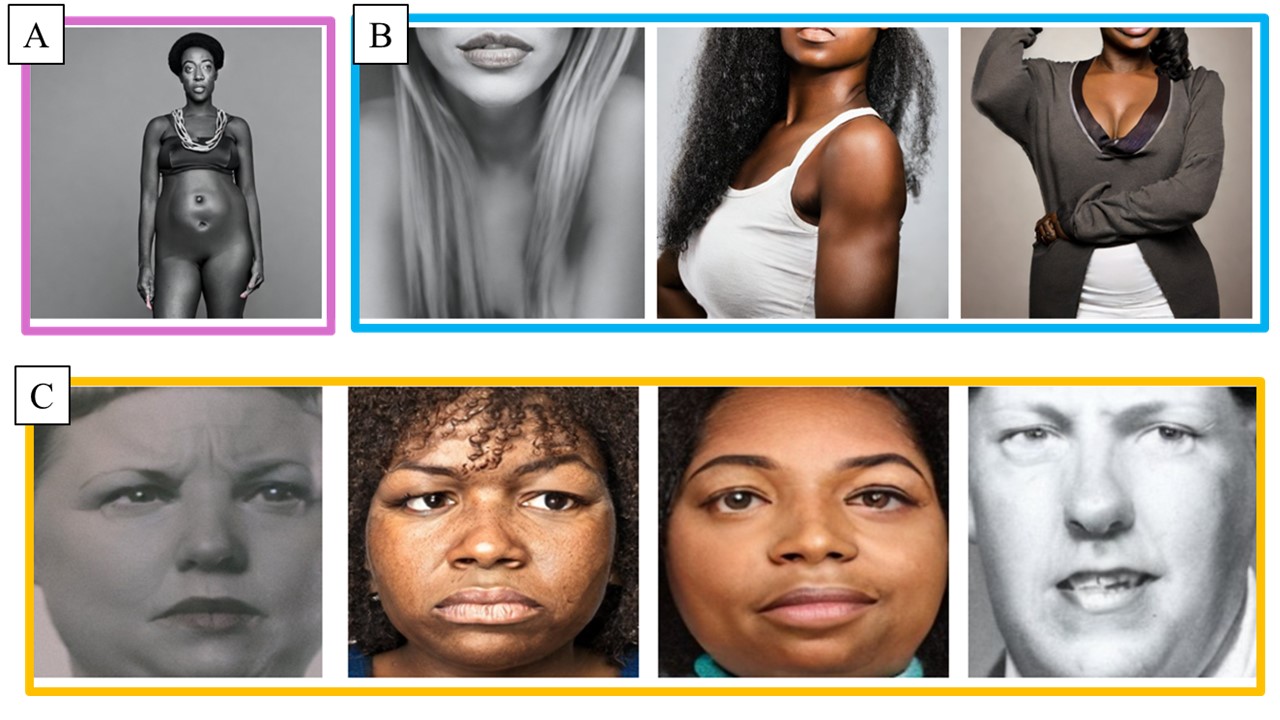}
    \caption{Controversial output for prompts referring to an unattractive black woman in (A) and (B) for attractive women (black/white). (B) Cases of "chubby" face in the unattractive groups}
    \label{fig:contro}
\end{figure}

Some controversial images were generated during the generation of images using Stable Diffusion. Despite the explicit prompt instructing the model to display a face, often the generated images represented bodies or body parts instead. One noteworthy observation is that body parts such as lips and prominently emphasized breasts were often generated for attractive women \ref{fig:contro}.B. Another remarkable case was the image generated in response to the prompt for unattractive Black women, as shown in figure \ref{fig:contro}.A. The image primarily depicts a partially nude, censored in intimate areas, pregnant abdomen. Additionally, there is a broader case of curvy faces among the non-attractive groups (figure \ref{fig:contro}.C). For men, this is also evident in the attribute detection results showing the attribute "chubby" among the top 10 detected. This last observation resonates with the stigmatization of fat bodies, which are often seen as unproductive and inefficient in Western culture \cite{harjunen2016neoliberal}. This connection again underscores societal biases' role in shaping AI outputs, where attributes such as 'chubby' linked to 'unattractive' become markers of negative judgment.

\section{A legal analysis of the essential requirements of gender classification systems’ operation
}
\label{sec:FeministPers}
\subsection{Data (e)quality under the scope of the AI Act and GDPR}
The analysis of the averageness theory revealed that specific physical attributes associated with attractiveness, such as facial features or perceived age, inadvertently influence gender classification within AI systems. Following the Gender Shades study \cite{buolamwini2018gender} hypothesis, it is possible that this result is linked to the type of descriptive features of the class most present in the training data.

The EU Non-discrimination legislation is crucial for safeguarding a high level of (e)quality in AI development and implementation settings. The obligation to respect the principle of non-discrimination is enshrined in EU primary law, in Article 2 of the Treaty on European Union (TEU)\footnote{ Treaty of the European Union, 1992, \url{https://www.cvce.eu/content/publication/2002/4/9/2c2f2b85-14bb-4488-9ded-13f3cd04de05/publishable_en.pdf}.}, Article 10 of the Treaty on the Functioning of the European Union (TFEU)\footnote{ Treaty on the Functioning of the European Union, 1958, \url{https://eur-lex.europa.eu/resource.html?uri=cellar:2bf140bf-a3f8-4ab2-b506-fd71826e6da6.0023.02/DOC_1&format=PDF}. } (requiring the Union to combat discrimination on several grounds) and Articles 20 and 21 of the EU Charter of Fundamental Rights (equality before the law and non-discrimination based on a non-exhaustive list of grounds)\footnote{ European Charter of Fundamental Rights,2012, \url{https://eur-lex.europa.eu/legal-content/EN/TXT/PDF/?uri=CELEX:12012P/TXT}.}. All prohibited grounds of discrimination, as listed in the Charter, are relevant regarding using algorithms.

In the AI context, algorithmic discrimination has become one of the critical points in the discussion about the consequences of an intensively \textit{datafied world} \cite{15_pena2019decentering}. The quality of the datasets used to train machine learning algorithms is of prime importance to the performance of AI systems, as “an algorithm is only as good as the data it works with” \cite{16_barocas2016big}. When data is gathered, it may contain socially constructed biases, inaccuracies or errors that must be tackled before any training based on this dataset \cite{17_europaEthicsGuidelines}. One of the reasons explaining the existence of bias in datasets is the “choice of subjects to the omission of certain characteristics or variables that properly capture the phenomenon we want to predict, to changes over time, place or situation, to the way training data is selected” \cite{18_dignum2021role}. AI algorithms trained on poor-quality information -both from the quantitative and qualitative point of view- can negatively affect the outputs or decisions of these mechanisms, leading to “incorrect model predictions” \cite{19_hacker2021legal}. It is worth recalling that a dataset “is always a reflection of the society from which the information has been obtained. If the society contains discriminatory elements and structures, these are also in the training data set” \cite{20_hoffmann2022training}. 

More than that, datasets used for training AI systems “may suffer from the inclusion of inadvertent historical bias, incompleteness and bad governance models”. The perpetuation of such biases could lead to inadvertent (in)direct prejudice and discrimination against certain groups or people, potentially exacerbating prejudice and marginalisation \cite{16_barocas2016big}. For instance, AI training datasets may exclude or under/misrepresent people from different geographical areas, neglecting or misconstruing their interests and needs. This exclusion can potentially exacerbate current inequalities and further marginalise these communities. \cite{21_hagerty2019global}. 

The AI Act mandates specific requirements and restrictions regarding the use of data for the development, training and testing of AI systems, encompassing factors like the quality, relevance, accuracy, representativeness and diversity of the data (Recitals 14a, 28a, 38, 43,44, 45 AI Act etc.), as well as the respect for the rights and interests of the data subjects and the data providers (Articles 9, 10, 54 and 55 AI Act). AI training and testing can be made either by datasets of real data or synthetic data (\textit{fake} data)- like in our case. Synthetic data is artificial data generated from original data (\textit{real} data) and a pre-trained model to reproduce the characteristics and structure of the original data (Recital 111 AI Act).

Whenever \textit{real} data is used, Articles 9 (1), 6 (1)(a), (b), and (f) GDPR are applied. Yet, the AI Act lacks explicit guidance on the proper procedures and legal basis for processing such data, particularly concerning consent acquisition and providing information and transparency, which are enshrined in the GDPR.  GDPR provides the proper legal framework by imposing different or additional conditions and constraints on the use of data for AI purposes. However, in practical terms, GDPR might be hardly invoked by the data subjects. The absence of this clarity raises questions about how the data subjects could ask to restrict processing (Article 18 GDPR) and to delete and erase data (Article 17 GDPR).  

At this point, we should highlight that the initial requirement outlined in Article 10 (5) (a) AI Act states that data processing under this article is permissible only when its goal, specifically bias detection and correction, \textit{"cannot effectively be achieved through processing synthetic or anonymised data."} This means synthetic or anonymised data should first be used to identify and correct bias. In contrast, real data can be used only when the requirement of synthetic or anonymised data is exhausted. Furthermore, when biases are based on sensitive data- like ethnicity data- the AI Act requires using anonymised data to process sensitive data as a primary bias detection and correction tool \cite{dominguez2022assessing}. AI-friendly data anonymisation tools align with the Recital 45 AI Act, which states, \textit{"Practices that are prohibited by Union law, including data protection law, non-discrimination law, consumer protection law, and competition law, should not be affected by this Regulation”}. It is worth mention\cite{dominguez2022assessing}ing that the inclusion of synthetic data in the AI Act was underlined by the European Commission’s Joint Research Centre \cite{MOSTLY_AI_2024}, supporting rebalancing mis/under-represented groups of people in ethnicity, gender, etc. 

\subsection{A gender-based risk assessment according to the AI Act and GDPR}

Another important aspect of gender classifiers, from a legal point of view, is to which extent they pose a risk to human rights (Articles 3 (2) and 27 AI Act). The AI Act adopts a risk-based approach (Article 9 AI Act) to oversee AI systems, categorising them into prohibited, high-risk and low-risk categories. However, the criteria and thresholds for determining the risk level of an AI system are not consistently clear. This lack of clarity may result in the exclusion of specific AI systems that may pose significant risks to data protection rights, such as those that process sensitive personal data or involve large-scale processing of personal data. For this purpose, a gender-based risk assessment in our study case is needed. 

To begin with, AI systems that profile individuals based on automated processing of personal data to assess various aspects of a person’s life, such as work performance, economic situation, health, preferences, interests, reliability, behaviour, location, or movement, are always considered high-risk AI systems. For instance, one area in which AI profiling is used is border management control by law enforcement agencies \cite{27_molnar2021robots,28_vavoula2021artificial}.

The classification rules for high-risk AI systems are enshrined in Article 6 (2) AI Act and Annex III AI Act. Remote biometric identification, biometric categorisation and emotion recognition systems are considered high-risk AI systems. Yet, the two requirements outlined in Article 6 (1) AI Act should be fulfilled to consider the above-mentioned AI systems as high-risk AI systems.  More particularly, the AI system should intended to be used as a safety component of a product, or the AI system is itself a product, covered by the Union harmonisation legislation listed in Annex I. Additionally, the product whose safety component under point (a) is the AI system, or the AI system itself as a product, is required to undergo a third-party conformity assessment, with a view to the placing on the market or the putting into service of that product according to the Union harmonisation legislation listed in Annex I.

Needless to say, these requirements are not fulfilled in our study case, which was an experiment for academic purposes (Article 2 (6) AI Act). Yet, for the above reasons, if this gender classifier was placed in the market, it would be considered a high-risk AI system. Moreover, solely for the legal analysis of our study case, we must stress that the data used is biometric derived from facial images. Biometric data is personal data resulting from specific technical processing relating to a natural person's physical, physiological or behavioural characteristics, which allow or confirm the unique identification of that person. Biometric data, like data revealing racial or ethnic origin or someone's sexual behaviour or sexual orientation, are considered a special category of data (Articles 3 (35) AI Act and  9 (1) GDPR). However, gender data is not. This difference is important, as GDPR does not offer extra protection for gender data processing, which signals that gender can be stored and processed without further constraints.

The legal analysis concludes with three key points. Firstly, it underscores the importance of enhancing data curation practices to improve the reliability and predictability of gender classifiers. This measure aims to mitigate the risk of generating biased outcomes that could result in arbitrary discrimination, unfair decisions, denial of services, or inappropriate interference with individuals' fundamental rights or freedoms. Data collection, curation and selection are essential components of AI risk management \cite{22_nistRiskManagement} and a fundamental aspect of the AI data governance framework. This framework aims to establish robust procedures ensuring high-quality data availability, labelling, and use. 

Secondly, the feminist theory of the social construction of gender offers insights into patriarchal power dynamics and the privileges of data (cis-normative) male domination. Through this lens, we can interrogate how to advance gender equality and prevent non-discrimination through data quality and curation. Given that "attractiveness" and beauty ideals differ across cultural and geographic boundaries, we endorse a feminist intersectional approach \cite{26_lutz2016framing} that acknowledges and accommodates these diversities. This approach aims to ensure that datasets used for AI training are not predominantly focused on white, middle-class, young, and heterosexual women.

Lastly, advocating for an interdisciplinary approach to developing AI applications becomes imperative in light of this background. This entails integrating the averageness theory from the field of psychology and analysing the legal safeguards to address gender bias in AI. The AI Act emphasises the necessity of such collaboration, as highlighted in Recital 142, to ensure that AI advancements yield socially advantageous results and address socio-economic disparities. This involves fostering cooperation among AI developers, experts in inequality and non-discrimination, academics, and other relevant stakeholders.

\section{Conclusion
}
\label{sec:conclusion}
This exploration of the complexities of gender perception and classification within artificial intelligence highlights the intricate interplay between technical capabilities and socio-psychological insights. Through the lens of the "averageness theory" and the creation of a synthetic dataset, the study illuminates the intrinsic biases associated with traditionally attractive features. Demonstrating how attractiveness, understood as youthfulness, the use of make-up, or particular facial expressions, algorithmically influences representations, particularly those of women.

A qualitative and quantitative analysis of the generated dataset reveals that men are also subject to gender stereotypes, with solid conformity to cis-normative stereotypes (for instance, having a beard or short hair). Consequently, the issue does not concern only the binary categorisation of gender but also how, within this dichotomy, stereotypes of femininity and masculinity, often tied to stereotypical beauty standards, are perpetuated and reinforced by AI algorithms. Such distortions can have unintended and harmful consequences for individuals and groups, underscoring the urgent need for ethical and responsible AI development.

In this context, the introduction of legislative discussions, particularly the efforts of the European Union exemplified by the AI Act, emphasises the necessity of integrating ethical, legal, and technical considerations to ensure the development of artificial intelligence technologies that respect human rights and equality. Indeed, from a legal standpoint, ensuring the regulation of diverse and representative datasets is fundamental \cite{pena2021oppressive}. However, adequate data curation requires a solid understanding of the various biases and discriminations that need attention and correction \cite{32_shams2023ai,33_fazelpour2022diversity}. For this reason, this work promotes the necessity of an interdisciplinary approach in the development phases of AI technologies. Beyond cognitive psychology, we highlight how studies such as the feminist theory of the social construction of gender offer insights into the dynamics of patriarchal power and the privileges of cis-normative male dominance in the data collection process.

Only through an integration of technological innovation, social analysis, and solid regulatory oversight can we achieve an AI that is not only advanced but also fair and inclusive. Artificial intelligence technologies that truly serve society require commitment from all social actors since implementing fairness measures within AI algorithms is indispensable despite their unavoidable non-neutrality \cite{33_fazelpour2022diversity}.
As we lay the groundwork for future explorations, this study advocates for a multidisciplinary strategy to identify and mitigate AI biases. The intertwining of technological innovation with robust regulatory oversight presents a promising avenue towards developing AI that is not only advanced but also aligned with the principles of fairness and inclusivity, thereby ensuring that AI serves the diverse needs of the global community.

\bibliography{references}

\end{document}